\documentclass[a4paper,12pt]{article}
\pdfoutput=1
\usepackage{hyperref}
\usepackage{epsfig}
\usepackage{amssymb}

\usepackage{setspace}

\usepackage{amsmath}
\usepackage{amssymb}
\usepackage{graphicx}
\usepackage{geometry}
\usepackage{subfigure}
\usepackage{url}

\newcommand{\bd}{\begin{displaymath}}
\newcommand{\ed}{\end{displaymath}}

\newcommand{\eg}{{\it e.g.}}

\newcommand{\be}{\begin{equation}}
\newcommand{\ee}{\end{equation}}
\newcommand{\br}{\begin{eqnarray}}
\newcommand{\bea}{\begin{eqnarray}}
\newcommand{\eea}{\end{eqnarray}}
\newcommand{\er}{\end{eqnarray}}
\newcommand{\ba}{\begin{array}}
\newcommand{\ea}{\end{array}}
\newcommand{\bi}{\begin{itemize}}
\newcommand{\ei}{\end{itemize}}
\newcommand{\bn}{\begin{enumerate}}
\newcommand{\en}{\end{enumerate}}
\newcommand{\bc}{\begin{center}}
\newcommand{\ec}{\end{center}}

\newcommand{\bs}{\begin{small}}
\newcommand{\es}{\end{small}}
\newcommand{\bfo}{\begin{footnotesize}}
\newcommand{\efo}{\end{footnotesize}}

 \def\({\left(}
 \def\){\right)}
 \def\[{\left[}
 \def\]{\right]}

\def\mHu{{m_{h_u}}}
 \def\mHd{{m_{h_d}}}
 \def\mHu2{{m_{h_u}^2}}
 \def\mHd2{{m_{h_d}^2}}
 \def\nn{\nonumber}
 
\def\haa{$H\to\gamma\gamma$}
\def\tqh{$q\; b \to t \,q' H$}

\def\haas{$H\to\gamma\gamma\;$}
\def\tqhs{$q\, b \to t \,q' H\;$}

\def\tqas{$q\, b \to t \,q' \gamma \gamma\;$}
\def\braasms{$BR^{SM}_{\gamma \gamma}\;$}
\def\braasm{$BR^{SM}_{\gamma \gamma}$}
\def\braas{$BR_{\gamma \gamma}\;$}
\def\braa{$BR_{\gamma \gamma}$}

\def\tqhpp{$p\, p \to t \,q\, H$}

\def\tqhspp{$p\, p \to t \,q\, H\;$}
\def\tqaaspp{$p\, p \to t \,q\, H\to t \,q\, \gamma \gamma\;$}
\def\tqas{$p\, p \to t \,q\, \gamma \gamma\;$}

\def\q2 {q^2}

\def\bt{\begin{table}}
\def\et{\end{table}}

\catcode`@=11 % This allows us to modify PLAIN macros.
\def \gsim{\mathrel{\mathpalette\@versim>}}
\def \lsim{\mathrel{\mathpalette\@versim<}}
\def \@versim#1#2{\lower0.4ex\vbox{\baselineskip\z@skip\lineskip\z@skip
     \lineskiplimit\z@\ialign{$\m@th#1\hfil##\hfil$%
     \crcr#2\crcr\sim\crcr}}}
\catcode`@=12 % at signs are no longer letters

\def\gappeq{\mathrel{\rlap {\raise.5ex\hbox{$>$}}
{\lower.5ex\hbox{$\sim$}}}}

\def\lappeq{\mathrel{\rlap{\raise.5ex\hbox{$<$}}
{\lower.5ex\hbox{$\sim$}}}}

\begin{document}
\pagestyle{empty}
%\begin{flushright}
%CERN-PH-TH-2012-078 \\
%\end{flushright}
%\vspace*{10mm}
\begin{center}
{\LARGE {\bf 
Single top and Higgs associated  production \\
\vspace{0.1cm}
as a probe of the $Ht\bar t$ coupling sign  \\
\vspace{0.2cm}
  at the LHC
}} \\
\vspace*{1.5cm}
{\large
 {\bf Sanjoy Biswas$^{a}$, Emidio Gabrielli$^{{b,c,}}$}\footnote{
On leave of absence from Dipart. di Fisica, Universit\`a di 
Trieste, Strada Costiera 11, I-34151 Trieste, Italy.},  {\bf and Barbara Mele$^{a}$}}
\vspace{0.3cm}

{\it
 (a) INFN, Sezione di Roma, \\ c/o Dipart. di Fisica, Universit\`a di Roma ``La Sapienza", \\ Piazzale Aldo Moro 2, I-00185 Rome, Italy}  \\[1mm]
{\it
 (b) NICPB, Ravala 10, Tallinn 10143, Estonia}  \\[1mm]
{\it
 (c) INFN, Sezione di Trieste, Via Valerio 2, I-34127 Trieste, Italy}  

\vspace*{2cm}{\bf ABSTRACT} \\
\end{center}
\vspace*{5mm}

\noindent
The LHC sensitivity to an anomalous Higgs coupling to the top quark  in the Higgs-top associated production is analyzed. Thanks to the strong destructive interference in the $t$-channel for standard model couplings, this process can be very sensitive to both the magnitude and the sign of a nonstandard top-Higgs coupling.
We analyze cross sections and the main irreducible backgrounds for the \haas decay channel. Sensitivities to an anomalous sign for the top Yukawa coupling are  found to be large. In particular,   at $\sqrt s=14$ TeV, assuming a universal rescaling in the Yukawa sector, a parton-level analysis with realistic selection cuts gives  a signal-to-background ratio $S/B\sim 5$, for $-1.5 \lappeq Y_t/Y_t^{ SM} 
\lappeq  0$. A number of events $S\simeq 10$ (with corresponding significances  
$\sim3\;\sigma$) are expected  for 60 fb$^{-1}$, to be compared with the standard-model   expectation $S\sim 0.3$.

\vfill\eject
%\pagestyle{empty}
%\clearpage\mbox{}\clearpage

\setcounter{page}{1}
\pagestyle{plain}

%INSERT YOUR TEXT HERE

\section{Introduction}
After many years of challenging experimental searches, a  signal consistent with a Higgs-boson has finally been consolidated at the LHC, with production rates compatible with the standard model ($SM$) predictions \cite{:2012gk,:2012gu}. We are now entering a new phase in  Higgs boson physics,  where (apart from looking for possible further Higgs physical states) the actual properties of this new particle will be determined by measuring its couplings  to the other known particles with  ever increasing precision.
The Higgs couplings to heavy vector bosons were indirectly detected through electroweak precision tests even before an Higgs signal direct observation~\cite{Baak:2012kk}. On the other hand, in order to constrain the Yukawa sector,  which describes the Higgs couplings to fermions,  we have to rely on the Higgs direct-observation profile.  Indeed, electroweak precision tests are  
not yet  sensitive at a measurable level to Yukawa coupling effects, which enter only at 2-loop level \cite{Gabrielli:2010cw}. There is now a first direct determination of the $H\to b\bar b$ decay recently claimed at Tevatron \cite{Aaltonen:2012qt}, while the LHC will likely  be sensitive at the $2\,\sigma$ level to both 
the $H b\bar b$ and  $H\tau\tau$ $SM$ couplings with the statistics
accumulated by the end of 2012 \cite{Murray}. Nevertheless, by making proper theoretical assumptions, 
one can already constrain at the LHC the actual characteristics of the Yukawa sector of a Higgs boson \cite{Plehn:2012iz}. For instance, by assuming  a {\it universal}  scale factor $C_f$ for the Higgs  Yukawa couplings
to all fermion species $f$
\be
Y_f = C_f \; Y_f^{SM}\, ,
\ee
(where {$Y_f^{SM}=m_f/{v}$} is the $SM$ Yukawa coupling and $v=\langle H \rangle$ is the vacuum expectation value of the Higgs 
field)
and a further scale factor describing the $HVV$ (where $V=W,Z$) couplings 
\be
g_{HVV} = C_V \; g_{HVV} ^{SM}\, ,
\ee
present data already constrains the fermion Yukawa couplings $Y_f$ to be inside two regions of values of opposite signs \cite{Azatov:2012bz,Espinosa:2012ir}. 
In particular,  if  no new physical degrees of freedom is present, the ATLAS fit pinpoints at 95\% C.L. the  intervals [-1.5,-0.5] and
[0.5,1.7] for  the scale factor $C_f$ , and the interval [0.7,1.4] for the $W/Z$ scale factor $C_V$ (where $C_{f,V}=$1 in the $SM$) \cite{atlas-1}.
On the other hand, the CMS fit restricts the analysis to positive values of the  Yukawa couplings,  and finds $C_{f}$ and
$C_{V}$ in the intervals  [0.3, 1.0] and
[0.7,1.2], respectively, at 95\% C.L.\footnote{Note that ATLAS and CMS obtain the 95\%C.L.
$C_{f}$ and $C_{V}$ intervals with different marginalization procedures.} \cite{cms-1}.

One should keep in mind  that an opposite sign in the Yukawa couplings  
with respect to the $SM$ prediction  would have a dramatic impact on the EW breaking
mechanism, even if 
its magnitude were close to 1. 
This is because the relative sign of the Higgs 
coupling to fermions and gauge vector bosons is crucial for recovering
the unitarity and renormalizability of the theory~\cite{Appelquist:1987cf}. Therefore, a 
negative sign in 
the Yukawa coupling would be an evidence of new physics that could manifest itself in many different ways.  Starting 
from the appearance of new Higgs bosons or weakly interacting resonances 
in the spectrum, in case one wants to recover perturbative unitarity, up to 
a new strongly interacting regime of weak gauge bosons with fermions above 
the TeV scale. Furthermore, 
flipping the sign of the $ttH$ coupling 
may lead to catastrophic vacuum instabilities~\cite{Reece:2012gi}.

 The only fermion species that sensibly contribute to the above  LHC fits are  the top quark (that enters in the loop of the main Higgs production mechanism $gg\to H$, and contributes with  maximal weight to the fits), the $b$ quark and the $\tau$ lepton, under the hypothesis  $C_{t}=C_{b}=C_{\tau}$.

The two non-degenerate  opposite-sign intervals for the top-Yukawa coupling arise from  
the $SM$ (destructive) interference between the $W$ vector-boson loop and the top-quark loop in the \haas amplitude. Indeed,  the present moderate enhancement observed 
in the \haas rate with respect to the $SM$ predictions (see \eg ~\cite{Corbett:2012dm,Giardino:2012dp}, and references therein) could be related to a decreased  top-Yukawa coupling, or even to  a change in the relative sign of the $W$-Higgs and top-Higgs couplings. The latter could considerably enhance the \haas branching ratio, without affecting the $gg\to H$ production rate.

A strictly fermiophobic Higgs interpretation, where $C_{f}=$0, and $C_{V}=$1, has been excluded 
by the LHC for the observed resonance.
 One should however keep in mind 
that, in {\it realistic} fermiophobic models, non-vanishing Yukawa couplings are generated at least at the radiative level by the chiral symmetry breaking induced by the non-vanishing fermion masses. Note that, in {\it effective} fermiophobic models, a  radiatively generated top  Yukawa coupling 
tends to have an opposite sign with respect to its $SM$ value \cite{Gabrielli:2010cw}.

In this paper, we address the problem of the determination of the relative sign of the $ttH$ and $WWH$ couplings through the study of the Higgs production in association with a single top quark at the LHC. While the magnitude of the top and $W$ couplings can be directly measured, respectively, through the  Higgs boson production in association with a top pair (see  \cite{Degrande:2012gr} for a recent study), and in vector-boson-fusion or $HW$-associated production, all these processes are not affected by the top and $W$ couplings relative sign. 
The $t$-channel  for single top and Higgs associated  production  is, on the other hand, particularly sensitive to the $Y_t$ and $g_{HWW}$ relative phase, because of the strong destructive interference in the $SM$ matrix element of the two Feynman graphs for  $q b \to t q' H$  in Figure 1  \cite{Tait:2000sh}.
%%%%%%%%%%%%%%%%%%%%%%
\begin{figure}[t]
\begin{center}
\includegraphics[width=0.5\textwidth]{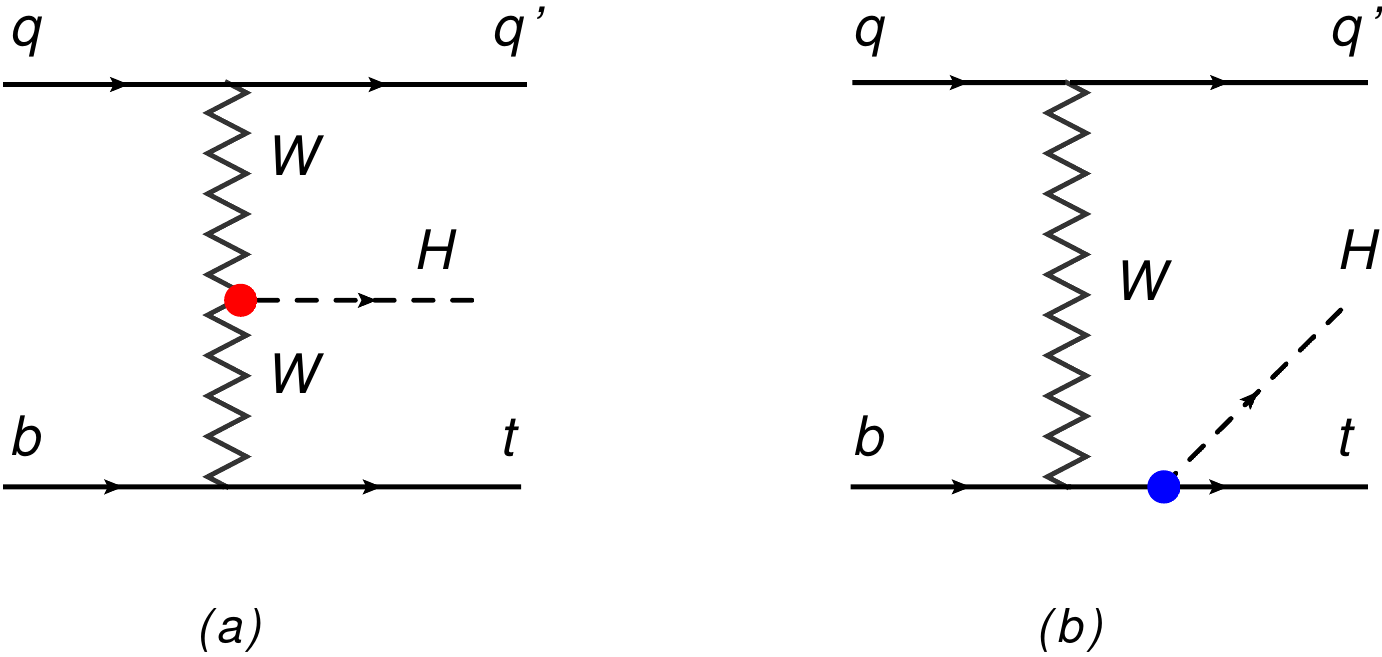}
\vspace{-0.3cm}
\caption{\small \it Feynman diagrams for the  single-top plus Higgs associated production 
in the $t$-channel \tqhs at  hadron colliders.  Higgs radiation by the initial $b$-quark line is not shown (see text). }
\label{fig:effvert}
\end{center}
\end{figure}
%%%%%%%%%%%%%%%%%%%%%%%%%

The associated production of a Higgs and a single top quark at the LHC and the SSC (the Superconducting Super Collider) was analyzed in the SM for a light Higgs boson in \cite{Stirling:1992fx}-\cite{Bordes:1992jy} 
(with \cite{Stirling:1992fx} focusing on the $H\rightarrow\gamma\gamma$ decay). In \cite{Maltoni:2001hu} $\,$ (see also \cite{Barger:2009ky}), a Higgs  decaying into $b\bar{b}$ pairs was studied for the same  
process at the LHC. However, quite negative conclusions were reached for both the $H\rightarrow\gamma\gamma$ and $H\rightarrow b\bar{b}$ modes in the SM.
The extension to a minimal supersymmetric two Higgs-doublet model slightly improves the expectations. In \cite{Barger:2009ky},  the $H\to WW^{(*)}, ZZ^{(*)}$ decays were studied for Higgs masses
150 GeV$<m_H< 200$ GeV. Both a $SM$ Higgs  and a variable $Y_t$-strength  setups (including the possibility of switching the $g_{HVV} ^{SM}$ sign) were considered. Good sensitivities in the above  
$m_H$ range were found for a non-standard relative sign of the Higgs couplings to $W$ and top quarks.

Among the three processes contributing to the  associated production of a Higgs and a single top quark process ($t$-channel $q\, b \to t q' H$, $s$-channel $q \bar q' \to t \bar b H$, and  $W$ associated production $g\, b \to W t H$),
here we concentrate on the $t$-channel $q\, b \to t q' H$. Indeed, as we will explicitly show, for a Higgs boson as light as 125 GeV, the latter has the largest cross section at the LHC, and  the highest sensitivity  to anomalous $Y_t$ couplings coming from the interference effects  of the two Feynman diagrams in Figure 1~\cite{Maltoni:2001hu}. We focus on the two-photon  decay \haa, and study the event rates for the signature corresponding to
\be
q\; b \to t \,q' H\to t \,q' \gamma \gamma\, ,
\ee
by applying realistic selection cuts based on present searches at the LHC. The rate suppression by the branching ratio ($BR$) for \haas  is then expected to be compensated by a better  
 signal-to-background event ratio $S/B$. We consider the hadronic $t\to b \,qq'$ decay, and  estimate the corresponding main {\it irreducible} backgrounds at  parton level, requiring the tagging of a $b$-jet in the final state. Present experimental studies of the \haas decay suggest that the contribution of the reducible backgrounds where 
photons are misidentified is  subdominant \cite{CMS-LAL,ATLAS-LAL}. We expect
a moderate contribution also from misidentified light and $b$ jets (see \eg \cite{Gabrielli:2007wf} as a relevant example).

We show that, while, for $Y_t$ close to its $SM$ value $C_{t}\sim1$, the  study of the  \haas decay channel in \tqhspp requires many hundreds of fb$^{-1}$ of integrated luminosity at 14 TeV, a  negative value of the top Yukawa coupling with $C_{t}\sim-1$
 would produce a detectable signal already with a few tens of fb$^{-1}$. A small but detectable number of  events (with excellent signal-to-background ratio $S/B$) are expected for $C_{t}\sim-1$ even at 8 TeV with the integrated luminosity available by the end of 2012.
 
 The plan of the paper is the following. In Section 2, we define the theoretical framework for the Higgs coupling dependence of the present study. In Section 3, we detail the top Yukawa dependence of the single-top plus Higgs associated production cross sections for the three main production mechanisms. We define the relevant backgrounds and selection cuts for the $t$-channel \tqhpp, in Section 4,
 where we also present results on signal ($S$) and background ($B$) event numbers. In Section 5, we discuss statistical significances of our results, and finally we conclude in Section 6.
\newpage 
 \section{Coupling parameter setups}

In the analysis of the potential of the channel \tqaaspp at the LHC, we will focus on the dependence 
on both magnitude and sign  of the $C_{t}$ scale factor.
Nevertheless, our results on the \tqhspp cross sections can straightforwardly be extended to a larger framework, where  the $W$ coupling factor $C_W$ has a non-standard value.
This follows from the  $C_{W}$ and $C_{t}$ dependence of the relevant production rates for the \tqhs process:
\be
d\,\sigma = d\,\sigma(C_{W} ,C_{t}) = |C_{W}|^2 \; d\,\sigma(1 ,C_{t}/C_{W})\,.
\ee
Hence, the critical parameter for cross sections in the present study is the relative phase of the  $C_{t}$ and $C_{W}$ scale factors, while a further variation in the $W$ coupling magnitude $|C_{W}|$  will just affect the production rate normalization.
From now on, we will then assume $C_W=C_V=1$.

Of course, the $C_{V}$ and $C_{f}$ setup have an impact not only on the Higgs production cross section but also on the branching ratio 
\be
BR_{\gamma\gamma}\equiv BR(H\to\gamma\gamma)
\ee
that enters the \tqas event rates. In order to make our results as model independent as possible, we will consider two different parameter setups :
\begin{itemize}
\item
{\it Universal Yukawa rescaling}, that is  assuming  just one free parameter  $C_f=C_t$  (and $C_V=1$) both in production and decay amplitudes. $BR_{\gamma\gamma}$ is then a function of  $C_t$, which enters  both the \haas width and the Higgs total width through $C_f$;
\item
{\it   $C_t$ and $BR_{\gamma\gamma}$ as independent parameters}  (and  $C_V=1$), with 
   $C_t$ affecting only  production  cross sections, and   $BR_{\gamma\gamma}$ describing the overall effect of new physics on the Higgs decay rate.
\end{itemize}
All the remaining couplings and physical degrees of freedom entering this study will be just the  $SM$ ones. 
The final  results for the  two setups can be easily related by just rescaling the event rates by the proper $BR_{\gamma\gamma}$ ratio.

\section{Signal production rates versus $C_t$}
In this section, we study the \tqhspp cross section dependence on the $C_t$ scaling factor,  assuming $C_V=1$. From now on, all the numerical cross sections discussed  will refer to the hadronic $p\,p$ collisions, even when the {\it partonic} initial state is shown. In order to compute the production rates at leading order, we used the MADGRAPH5 (v1.3.33) software package \cite{Alwall:2011uj}, with the CTEQ6L1 
 parton distribution 
functions \cite{CTEQ6L1}. We set both the  factorization and renormalization scales at the value  $Q=\frac{1}{2}(m_H+m_t)$ for the  \tqhspp 
signal, where $m_t$ is the top-quark mass. The other relevant parameters entering our computation  are set as follow
\cite{:2012gk,:2012gu,Aaltonen:2012ra,Beringer:1900zz}:
\bea
m_H&=&125 {\rm ~GeV},\;\;\;\;\;\;\;\;\;\;~m_t=173.2 {\rm ~GeV}, \nn\\
M_Z&=&91.188 {\rm ~GeV},\;\;\;\;~M_W=80.419 {\rm ~GeV}, \nn\\
m_b&=&4.7 {\rm ~GeV},\;\;\;\;~{\rm and} \;\;~\alpha_S(M_Z)=0.118\;. \nn
\eea
The $SM$   \haas branching ratio \braasms   was obtained by HDECAY~\cite{Djouadi:1997yw}, while the model dependent \braas versus $C_f$ has been evaluated 
via the leading-order $H$ partial widths~\cite{Djouadi:2005gi}, improved by normalizing the result by a factor \braasm/$BR_{\gamma \gamma}^{C_f=1}$ (where $BR_{\gamma \gamma}^{C_f=1}$ is the leading-order evaluation of the $SM$ branching ratio). For reference in the following discussion, the relevant $SM$ cross sections $\sigma$ and \braas are (summing up cross sections over the two charge-conjugated channels)\footnote{The contribution to the \tqhspp cross section of the amplitude where the  Higgs is radiated by the initial $b$-quark line is small (at the per-mil level in the $C_t$ range relevant here),  and will be neglected in the present analysis.}
\bea
\sigma(q\, b \to t \,q' H)^{SM}&\simeq& 15.2\; {\rm fb\;\;\;\;\;\;\;\;\;\; at} \;\;\;\sqrt s = 8\; {\rm TeV}\\
\sigma(q\, b \to t \,q' H)^{SM}&\simeq& 71.8\; {\rm fb\;\;\;\;\;\;\;\;\;\; at} \;\;\;\sqrt s = 14\; {\rm TeV}
 \\
\nn \\
BR^{SM}_{\gamma \gamma}&\simeq& 2.29\cdot 10^{-3} 
\eea
%%%%%%%%%%%%%%%%%%%%%%%%%%%%...Figure-1
\begin{figure}[t]
\begin{center}
\includegraphics[width=0.6\textwidth]{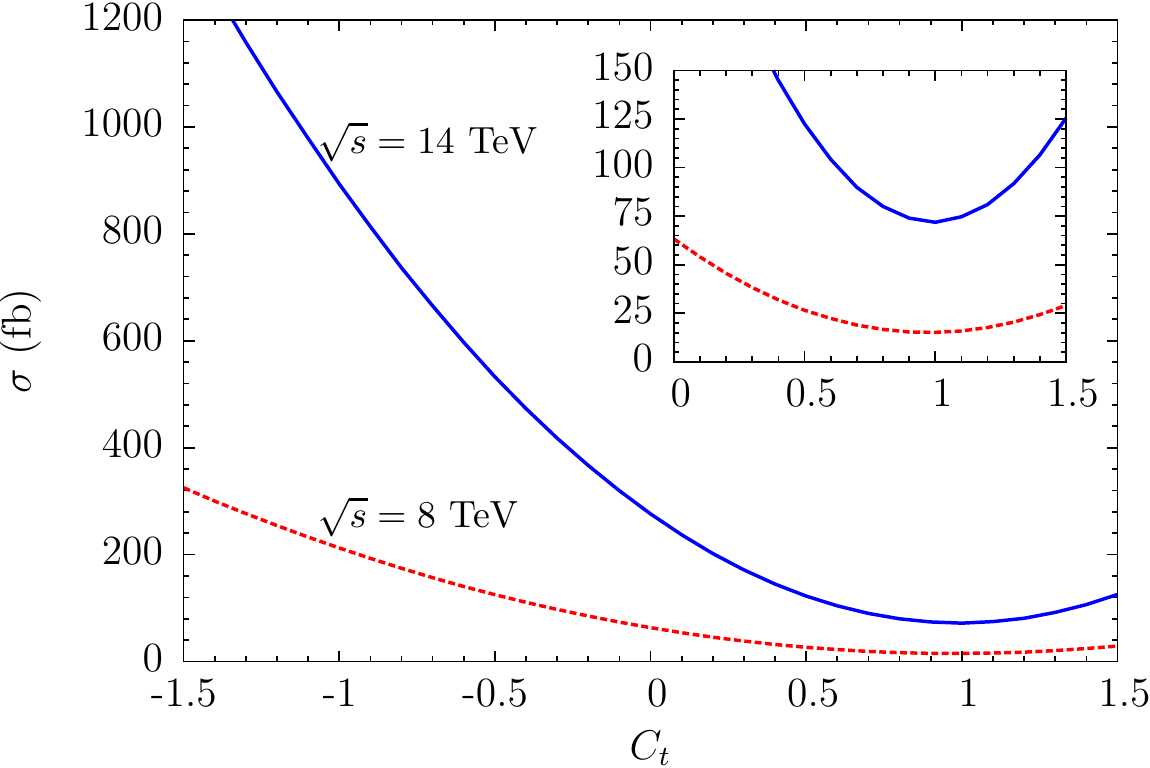}
\vspace{-0.3cm}
\caption{\small \it Production cross sections  for \tqhspp versus $C_t$, for $\sqrt s=8 {\rm ~and} ~14$ TeV. The inside plot is an enlargement of the positive $C_t$ region. }
\label{fig:xsec}
\end{center}
\end{figure}
%%%%%%%%%%%%%%%%%%%%%%%%%%%%...
In Figure~\ref{fig:xsec}, we plot the \tqhspp production cross-section versus $C_t$, for $\sqrt s=$ 8 TeV and 14 TeV. Throughout this work we focus on the range  
\be
-1.5<C_t<1.5\; ,
\ee
where the $C_t$ dependence is more critical, and  the most favored regions of the LHC  
fits  lie~\cite{atlas-1,cms-1}. Figure~\ref{fig:xsec} $\,$ shows that in the $SM$ $C_t=1$ case  
the destructive effect of the interference of the two diagrams in Figure~\ref{fig:effvert} is maximal, and that a sign change in  $Y_t$ produces a dramatic enhancement in the \tqhspp production cross sections.

Similarly, the destructive interference between the $W$ and top loops in the \haas decay  gives rise to an enhancement in the width $\Gamma_{\gamma \gamma}\;$ after switching the $C_t$ sign. On the other hand, the  overall \braas dependence on $C_t$ is mostly influenced, in the 
$C_f=C_t$ hypothesis, by the $C_f$ impact on the Higgs dominant decay widths into $b$ quarks, and $\tau$ leptons.  

Since the cross section and \braas dependencies on $C_t$ are both crucial for the results of the
%%%%%%%%%%%%%%%%%%%%%%%%%%%%...Figure-3
\begin{figure}[thp]
\begin{center}
%\vspace*{-2.2cm}
\includegraphics[width=0.6\textwidth]{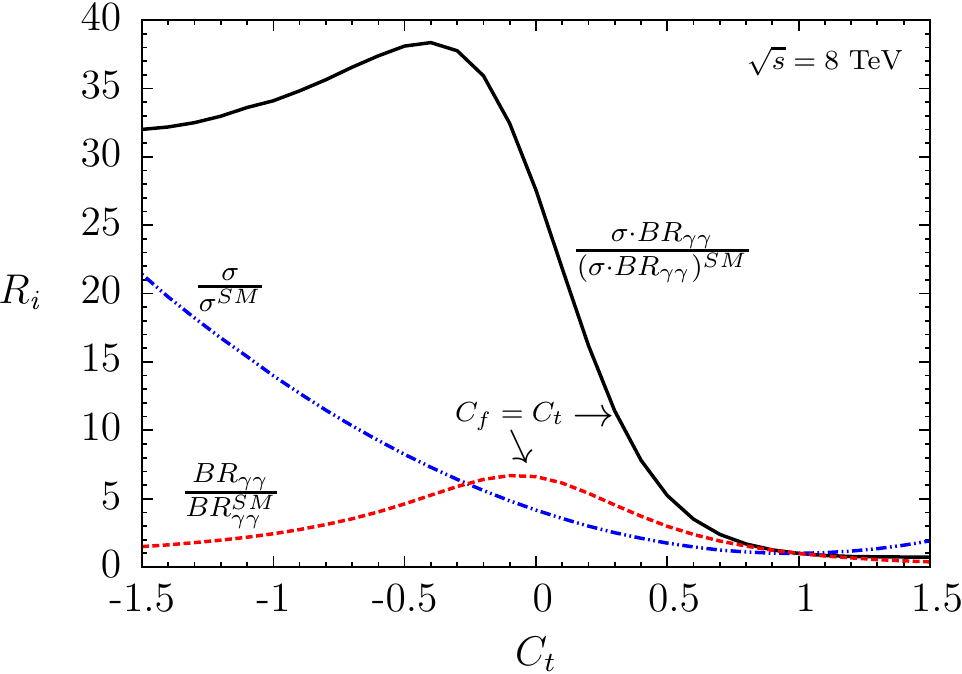}
\vskip 10pt
\includegraphics[width=0.6\textwidth]{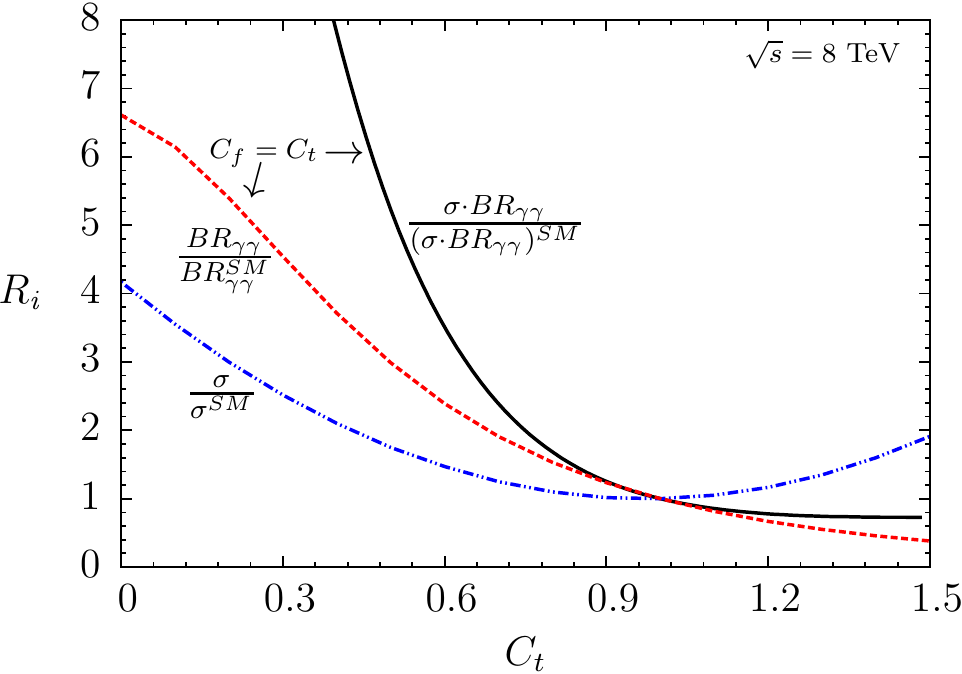}
\vspace{-0.3cm}
%\vspace*{1cm}
\caption{\small \it Enhancement factors $R_i$ for the \tqhspp production cross section 
$\sigma$,
\braa, and  their product with respect to their $SM$ values, versus $C_t$, 
  for $\sqrt s=8$ TeV. The lower plot is just an enlarged view of the positive $C_t$ range.} 
\label{fig:ratio-8}
\end{center}
\end{figure}
%\vspace{-1.0cm}
%%%%%%%%%%%%%%%%%%%%%%%%%%%%...Figure-3
\begin{figure}[thp]
\begin{center}
%\vspace*{-2.2cm}
\includegraphics[width=0.6\textwidth]{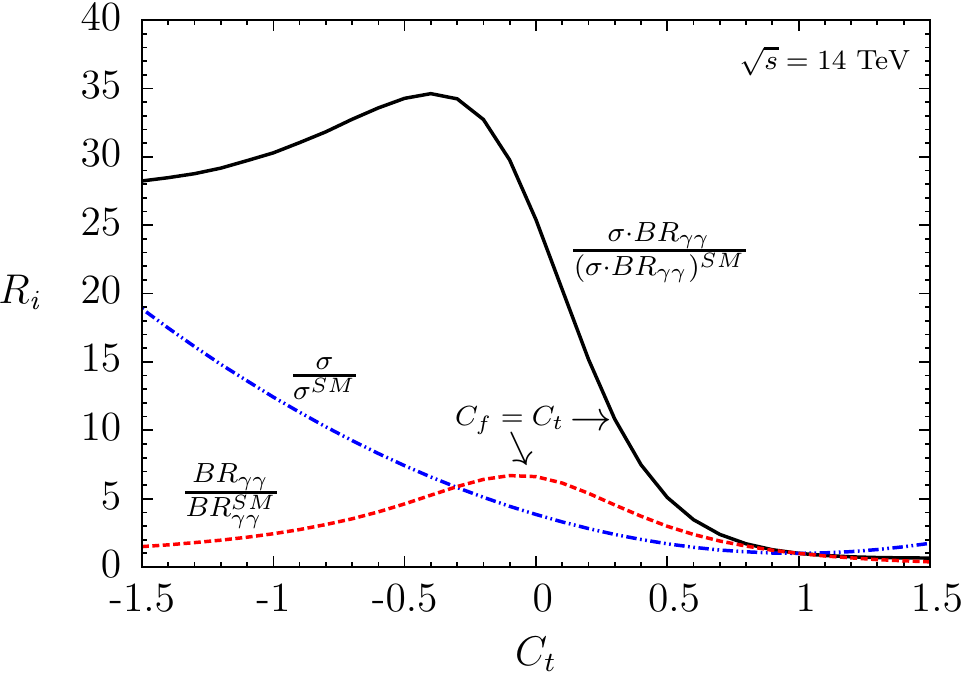}
\vskip 10pt
\includegraphics[width=0.6\textwidth]{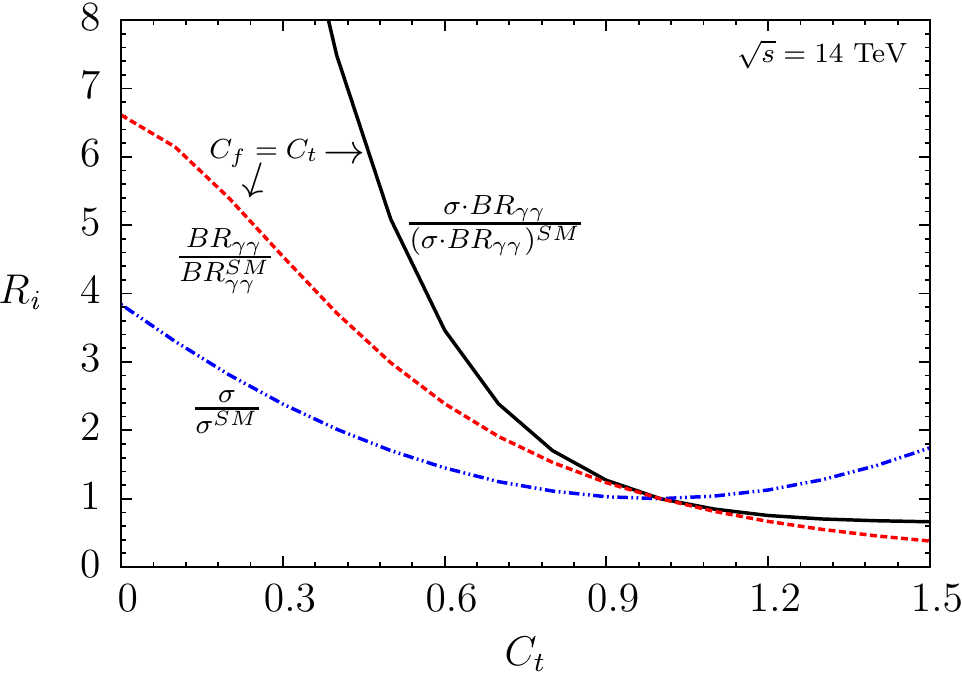}
\vspace{-0.3cm}
%\vspace*{1cm}
\caption{\small \it Enhancement factors $R_i$ for the \tqhspp production cross section 
$\sigma$,
\braa, and  their product with respect to their $SM$ values, versus $C_t$, 
  for $\sqrt s=14$ TeV. The lower plot is just an enlarged view of the positive $C_t$ range.} 
\label{fig:ratio-14}
\end{center}
\end{figure}
%%%%%%%%%%%%%%%%%%%%%
 present analysis, we plot in Figure~\ref{fig:ratio-8} (for $\sqrt s=8$ TeV) and 
Figure~\ref{fig:ratio-14} (for $\sqrt s=14$ TeV) the ratios $R_i$ of the $C_t$ dependent 
$\sigma$(\tqhpp), \braa, and product $\sigma$(\tqhpp)$\cdot$\braas over the corresponding $SM$ values, 
for $-1.5<C_t<1.5$. An enlargement of the positive $C_t$ range  is given in the lower plots of both figures. Going to negative  $C_f$ values has a dramatic effect on both cross sections and production rates for \haa. On the other hand, \braas is mostly sensitive to a reduction of the $|C_f|$ magnitude, and less influenced by the $C_f$ sign. 

For the sake of completeness, we also evaluated the total cross section and $C_t$ dependence  
for the top-Higgs associated production with a  $W$ in the process  $g\, b \to W t H$,
and for the $s$-channel $q \bar q' \to t \bar b H$.
We obtain (summing up  cross sections over the two charge-conjugated channels),
at $\sqrt s =$ 14 TeV,
\bea
\sigma(g\, b \to W t H)^{SM}\;\;\;\;&\simeq& 16.0\; {\rm fb}\, ,
 \\
\sigma(g\, b \to W t H)^{C_t=0}\;\;&\simeq& 34.9\; {\rm fb}\, ,
  \\
 \sigma(g\, b \to W t H)^{C_t=-1}&\simeq& 139.\; {\rm fb}\, ,
   \\
\nn \\
\sigma(q \,\bar q' \to t \, b \,H)^{SM}\;\;\;\;&\simeq& 2.26\; {\rm fb}\, ,
 \\
\sigma(q \,\bar q' \to t \, b\, H)^{C_t=0}\;\;&\simeq& 1.49\; {\rm fb}\, ,
  \\
 \sigma(q \,\bar q' \to t \, b \,H)^{C_t=-1}&\simeq& 0.39\; {\rm fb}\; , 
\eea
to be compared with the $t$-channel cross sections, at  $\sqrt s =$ 14 TeV,
\bea
\sigma(q\, b \to t \,q' H)^{SM}\;\;\;\;&\simeq& 71.8\; {\rm fb}\, ,
 \\
\sigma(q\, b \to t \,q' H)^{C_t=0}\;\;&\simeq& 276.\; {\rm fb}\, ,
  \\
 \sigma(q\, b \to t \,q' H)^{C_t=-1}&\simeq& 893.\; {\rm fb}\; .
 \eea
Although there is a nice sensitivity to $C_t$ also in the $W$-associated production, we do not concentrate on this process here, because of its lower rates with respect to the 
$t$-channel \tqh.  Nevertheless, we checked that its contribution to our 
event selection analysis, optimized for the \tqhspp process, is negligible.

\section{Signal versus irreducible backgrounds}

The  irreducible $SM$ backgrounds for the \tqaaspp process, for the top hadronic decay $t\to b \,q\,q'$, correspond to final states containing  two photons, one $b$-jet,  and at least three light jets, i.e., $2\,\gamma+b\,+(\geq\;3\,j)$.
The main partonic channels contributing are top production  (either single or in pair) and multi-jet final states, when accompanied by two high-$p_T$ photons, 
\bea
pp&\to& 2\,\gamma+t+j\,,   \\
pp&\to& 2\,\gamma+\bar{t}\,t\,,   \\
pp&\to& 2\,\gamma+b+3\,j \,,
\eea
with subsequent decay  $t\to b \,q\,q'$. We always require the $b$-jet identification in the final state.

To study the above channels we have used the same simulation package and parton distribution functions described in Section~3 with the renormalization and
factorization scale set at 
the default dynamical scale value in MADGRAPH5~\cite{Alwall:2011uj}.

As discussed in Section~1, we postpone the study of the channels contributing to the {\it reducible} background through misidentified particles to a more in-depth analysis, being confident that the bulk of the final background will originate from the irreducible one. 
In our analysis,  
 jets are defined at the parton level.

%%%%%%%%%%%%%%%%%%%%%%%

In order to tag an event, we require the final particles to pass the following 
 selection criteria, modeled according to present searches~\cite{CMS-LAL,ATLAS-LAL} :
\begin{eqnarray}
\begin{tabular}{l}
$p_{T}^{\gamma_1}>40$~GeV, \qquad $p_{T}^{\gamma_2}>30$~GeV, \qquad $p_{T}^{j,b}>25$~GeV, 
\qquad $|\eta_{\gamma,b}|<2.5$,  \qquad $|\eta_{j}|< 4.5$.
\end{tabular}
\label{tag1}
\end{eqnarray}
From now on, $b$ stands for a $b$-jet. We assume a $b$-tagging efficiency of 60\%,
which should guarantee a very good light-jet rejection factor~\cite{atlas-btag}. 
  The isolation requirement between the final state
photons, light jets $j$, and $b$-jets is 
\begin{equation}
\Delta R_{i,j}=\sqrt{ \Delta \eta^2_{i,j}+\Delta \phi^2_{i,j}}> 0.4\;,
\label{tag2}
\end{equation}
with {\it i} and  {\it j} running over all the final photons and jets (including $b$-jets),
$ \Delta \eta$ is the rapidity gap, and $ \Delta \phi$ is the azimuthal angle gap between the 
particle pair.
Because the \tqhspp signal comes from a t-channel $W$ exchange process, the light jet in the final state  
has normally large rapidity and high transverse momentum. In the chain of subsequent cuts applied, we therefore first require a forward jet (defined as the
highest rapidity light-jet in the final state) with $|\eta|>2.5$ and $p_T> 30 \;(50)$  GeV at $\sqrt s=$ 8 TeV (14 TeV)~\cite{Tackmann,Barger:2009ky}. Then, we  require a top quark fully reconstructed in the hadronic mode, i.e., the invariant mass
of 3-jets (out of which one is a $b$-jet) must peak at  the top mass within a mass window of 20 GeV.
Then, the invariant mass of the two light jets, contributing to the top quark invariant
mass, must peak at the $W$ 
mass within a mass window of 15 GeV. Finally, we impose that  
the invariant mass of
 the di-photon system reconstructs  
the Higgs mass centered at 125 GeV within a mass window of $\pm 3$ GeV.  
This set of  selection cuts is quite conservative, and consistent with the present experimental analyses at the LHC.

%\begin{scriptsize}
%%%%%%%%%%%%%%%%%%%%%%%%%%%%%%%%%
\begin{table}[thp]
\begin{center}
\begin{tabular}{||c||c|c|c||c|c|c|c||}
\hline 
\hline
$\sqrt{s}=8$ TeV \bs(60 fb$^{-1}$) \es & \multicolumn{3}{c||}{{\it Signal (S)}} & \multicolumn{4}{c||}{{\it Background (B)}} \\
%\cline{2-8}
\hline
{\it Cut}  & \bs {\bf $C_t=-1$}\es &  \bs{\bf $C_t=0$}\es & 
\bs{\bf $C_t=1$}\es & \bs$2\gamma+t+j$\es & \bs$2\gamma+t\bar{t}$\es &  
\bs$2\gamma+b+3j$\es& {\it \bs $B_{tot}$\es} \\
\hline
\bs$2\gamma+b+ (\geq  \;$3\,j) \es&  7.7  &  6.1  &  0.21  & 9.8 & 11 & 299  & 320 \\
\bs$|\eta^F_{j}|>2.5$ \& $p_{T}^{F}>30$~GeV \es& 3.7  & 3.0 & 0.09 & 4.0 & 0.46  & 26  & 31 \\
\bs$|M_{bjj}-m_{t}|<20$ GeV\es & 3.6  &  2.9   & 0.09  & 4.0 & 0.29  & 6.5 & 11 \\
\bs$|M_{jj(top)}-M_{W}|<15$ GeV \es& 3.4  &  2.8  & 0.08  & 3.8  & 0.23  & 2.1 & 6.1 \\
\bs$|M_{\gamma\gamma}-m_H|<3$ GeV\es & 3.4  &  2.8  & 0.08 & 0.14 & 0.02 & 0.68  & 0.84 \\
\hline
\hline
\end{tabular}
\caption{\small \it Number of events passing sequential cuts for the signal \tqaaspp and irreducible $SM$ backgrounds  at $\sqrt s=8$ TeV, and integrated luminosity of $60$ fb$^{-1}$, assuming $C_f=C_t$.} 
\label{tab:eventrate-8}
\vspace*{-0.3cm}
\end{center}
\end{table}          
%>>>>>>>>>>>>>>>>>>>>>>>>>>>>>>>>>>>>>>>>>>>>>>>>>>>>>>>>>>>>>>>>>>>>>>>>>>>>>>>>>>>>>>>>>>>>>>>>
\begin{table}[hpt]
\begin{center}
\begin{tabular}{||c||c|c|c||c|c|c|c||}
\hline 
\hline
 $\sqrt{s}=14$ TeV  \bs(600 fb$^{-1}$) \es & \multicolumn{3}{c||}{{\it Signal (S)}} & \multicolumn{4}{c||}{{\it Background (B)}} \\
%\cline{2-8}
\hline
 {\it Cut}  & \bs {\bf $C_t=-1$}\es&  \bs{\bf $C_t=0$}\es & 
\bs{\bf $C_t=1$}\es & \bs$2\gamma+t+j$\es & \bs$2\gamma+t\bar{t}$\es &  
\bs$2\gamma+b+3j$\es & {\it \bs $B_{tot}$\es} \\
\hline
\bs$2\gamma+b+ (\geq  \;$3\,j) \es& 311  & 249  &  8.9 &  407 &  396  & 12079 & 12881 \\
\bs$|\eta^F_{j}|>2.5$ \& $p_{T}^{F}>50$~GeV \es& 121  &  99  & 3.5 &  161 & 19 & 551 & 731 \\
\bs$|M_{bjj}-m_{t}|<20$ GeV \es& 118   & 97  &  3.5 & 161 & 11 & 136 & 308 \\
\bs$|M_{jj(top)}-M_{W}|<15$ GeV\es & 112   & 92 & 3.3 &  154 & 8.3 & 43  & 205 \\
\bs$|M_{\gamma\gamma}-m_H|<3$ GeV\es & 112  & 92 &  3.3 &  7.2  & 0.28 & 14 & 22 \\
\hline
\hline
\end{tabular}
\caption{\small \it Number of events passing sequential cuts for the signal \tqaaspp and irreducible $SM$ backgrounds  at $\sqrt s=14$ TeV, and integrated luminosity of $600$ fb$^{-1}$, assuming $C_f=C_t$.} 
\label{tab:eventrate-14}
\end{center}
\end{table}         
\noindent

The results of the above selection procedure are shown in Table~\ref{tab:eventrate-8} 
(for $\sqrt s=8$ TeV, and integrated luminosity of 60 fb$^{-1}$) and
Table~\ref{tab:eventrate-14} (for $\sqrt s=14$ TeV, and integrated luminosity of 600 fb$^{-1}$) .
At $\sqrt s=8$ TeV,  60 fb$^{-1}$ could correspond to the  maximal  integrated
luminosity expected by collecting both the  ATLAS and CMS data by the end of 2012.

The numbers of events passing the sequential cuts  defined above are reported 
for the \tqaaspp signal, for different $(C_t=\pm 1,0)$ values (assuming $C_f=C_t$ in \braa), and main irreducible backgrounds.
The first row refers to the total number of events passing the photon- and jet-tagging definition
in Eqs.~(\ref{tag1}) and (\ref{tag2}), while the last column shows  the total number of background events $B_{tot}$.
One can  see the efficiency of the different cuts applied to enhance the signal-to-background ratio. The signal rate is only affected by the forward-jet tag, with a corresponding reduction of a factor about 2, and passes almost unaltered the remaining cuts (the small reduction in the  event numbers arising from the light-jets originating from the $t$ decay being tagged as forward jets). The largest contribution to the background comes from the $2\,\gamma+b+3\,j \,$ non-resonant final state, that is considerably affected by both the forward-jet cut and the top- and Higgs-resonance requirements. The next background for importance is the 
single-top production $2\,\gamma+t+j$, while the top-pair channel $2\,\gamma+\bar{t}\,t\,$ contributes to $B_{tot}$ negligibly.

%%%%%%%%%%%%
%%%%%%%%%%%%%%%%%%%%%%%%%%%%%%%%%
\begin{table}[tp]
\begin{center}
\begin{tabular}{||l||c|c|c|c|c|c|c||}
\hline 
\hline
\bs $\sqrt{s}=14$ TeV (600 fb$^{-1}$) \es  &  \begin{footnotesize}{\bf $C_t=-1.5$}\end{footnotesize}  & \begin{footnotesize}{\bf $C_t=-1.$}\end{footnotesize} & 
\begin{footnotesize}{\bf $C_t=-0.5$}\end{footnotesize}  & \begin{footnotesize}{\bf $C_t=0$}\end{footnotesize} & 
\begin{footnotesize}{\bf $C_t=0.5$}\end{footnotesize} & \begin{footnotesize}{\bf $C_t=1.$}\end{footnotesize} & \begin{footnotesize}{\bf $C_t=1.5$}\end{footnotesize} \\
\hline
$S\,[$\bs$\,2\gamma+b+ (\geq  \,3\,j)\,]_{tag}$ \es & 287 & 311 & 338  & 249 & 47 & 8.9 &  6.6
 \\
$\;\;\;\;S\,${\small[{\it passing cuts}]} & 104  & 112 & 122  & 92  &  17 &  3.3 &  2.2\\
\hline
\bs$\;\;\;\;\;\;\;\;{S}/{\sqrt{S+B}}$ \es& 9.3 & 9.7  & 10.  & 8.6  & 2.7  & 0.67  & 0.45 \\
\hline
\hline
\end{tabular}
\caption{\small \it Number of tagged events, and number of events passing all selection cuts for the \tqaaspp signal  at $\sqrt s=14$ TeV, and integrated luminosity of $600$ fb$^{-1}$, versus $C_t\;$(assuming $C_f=C_t$). Statistical significances of the signal are shown, based on the background event numbers presented in Table~\ref{tab:eventrate-14}.} 
\label{tab:sig-14}
\end{center}
\end{table}   
%\vspace*{-0.3cm}      
%%%%%%%%%%%%%%%%%%%%%%%%%....Comment out this table later.
\begin{table}[t]
\begin{center}
\begin{tabular}{||l||c|c|c|c|c|c|c||}
\hline 
\hline
 \bs $\sqrt{s}=14$ TeV (60 fb$^{-1}$) \es &  \bfo{\bf $C_t=-1.5$}\efo  & \bfo{\bf $C_t=-1.$}\efo & 
\bfo{\bf $C_t=-0.5$}\efo  & \bfo{\bf $C_t=0$}\efo & 
\bfo{\bf $C_t=0.5$}\efo & \bfo{\bf $C_t=1.$}\efo & \bfo{\bf $C_t=1.5$}\efo \\
\hline
$S\,[$\bs$\,2\gamma+b+ (\geq  \,3\,j)\,]_{tag}$ \es & 29 & 31 & 34  & 25 & 4.7 & 0.89 &  0.66 \\
$\;\;\;\;S\,${\small[{\it passing cuts}]} & 10  & 11 & 12  & 9.2  &  1.7 &  0.33 &  0.22\\
\hline
\bs$\;\;\;\;\;\;\;\;{S}/{\sqrt{S+B}}$ \es& 2.9 &  3.1 & 3.2 & 2.7 & 0.86 & 0.21  & 0.14  \\
\hline
\hline
\end{tabular}
\caption{\small \it Number of tagged events, and number of events passing all selection cuts for the \tqaaspp signal  at $\sqrt s=14$ TeV, and integrated luminosity of $60$ fb$^{-1}$, versus $C_t\;$(assuming $C_f=C_t$). Statistical significances of the signal are shown, based on the background (rescaled) event numbers presented in Table~\ref{tab:eventrate-14}.} 
\label{tab:sig-14bis}
\end{center}
\end{table}   
%\vspace*{-0.3cm}   
%\vspace*{0.8cm}            
%%%%%%%%%%%%%%%%%%%%%%%%%%%%%%%%%%%
\begin{table}[hb]
\begin{center}
\begin{tabular}{||l||c|c|c|c|c|c|c||}
\hline 
\hline
\bs $\sqrt{s}=8$ TeV (60 fb$^{-1}$) \es &  \begin{footnotesize}{\bf $C_t=-1.5$}\end{footnotesize}  & \begin{footnotesize}{\bf $C_t=-1.$}\end{footnotesize} & 
\begin{footnotesize}{\bf $C_t=-0.5$}\end{footnotesize}  & \begin{footnotesize}{\bf $C_t=0$}\end{footnotesize} & 
\begin{footnotesize}{\bf $C_t=0.5$}\end{footnotesize} & \begin{footnotesize}{\bf $C_t=1.$}\end{footnotesize} & \begin{footnotesize}{\bf $C_t=1.5$}\end{footnotesize} \\
\hline
$S\,[$\bs$\,2\gamma+b+ (\geq  \,3\,j)\,]_{tag}$ \es & 7.4  &  7.7  & 8.7  & 6.1  & 1.1 & 0.21 &  0.16 \\
$\;\;\;\;S\,${\small[{\it passing cuts}]} & 3.2  &  3.4   & 3.8  & 2.75  & 0.52 & 0.08 & 0.06 \\
\hline
\bs$\;\;\;\;\;\;\;\;{S}/{\sqrt{S+B}}$ \es& 1.6  & 1.7  & 1.8 & 1.5 & 0.45  &  0.08 & 0.06 \\
\hline
\hline
\end{tabular}
\caption{\small \it Number of tagged events, and number of events passing all selection cuts for the \tqaaspp signal  at $\sqrt s=8$ TeV, and integrated luminosity of $60$ fb$^{-1}$, versus $C_t\;$(assuming $C_f=C_t$). Statistical significances of the signal are shown, based on the background event numbers presented in Table~\ref{tab:eventrate-8}.} 
\label{tab:sig-8}
\end{center}
\end{table}         
%\noindent 

\section{Signal significance versus $C_t$}
%%%%%%%%%%%%%%%%%%%%%%%%%%%%%%%%%%%%%%%%%%%%%%%%%%%%%%%%%%%%
By comparing the signal- and background- event numbers passing all the selection cuts in the last row  of 
Table~\ref{tab:eventrate-8}  and 
\ref{tab:eventrate-14}, the sensitivity of the \tqaaspp process to a change of sign  of $Y_f$
gets clearly manifest. 
While a $SM$ coupling configuration $Y_f/Y_f^{SM}\simeq 1$ provides a signal-to-background ratio
$S/B\sim 10\% \,(15\%)$ at $\sqrt s=8\, (14)$ TeV, 
when $Y_f/Y_f^{SM}\simeq-1$ one reaches 
$S/B$ as large as $\sim 4\, (5)$ at $\sqrt s=8\,(14)$ TeV.
At $\sqrt s=8$ TeV, the signal event number is quite small. For $C_f\simeq -1$, one expects
about $S=3$  with $B\lappeq 1$, for  60 fb$^{-1}$.

On the other hand, at $\sqrt s=14$ TeV with  600 fb$^{-1}$, 
one gets about $S\simeq 100$  with $B\simeq 20$  in all the negative $Y_f$  range considered $(-1.5\lappeq C_f \lappeq 0)$, with corresponding statistical significances 
$S/\sqrt{(S+B)}\sim 10$.  This  is shown  in Table~\ref{tab:sig-14}, where,  
for a set of different $C_t$ values, we report  the corresponding 
significances $S/\sqrt{(S+B)}$ at $\sqrt s=14$ TeV, and integrated luminosity of $600$ fb$^{-1}$.
For reference, we also show the number of tagged signal events according to the object definitions in  Eqs.~(\ref{tag1}) and (\ref{tag2}), and the number of tagged events passing all the  sequence of selection cuts. The relevant number of background events ($B=B_{tot}$)  can be found in Table~\ref{tab:eventrate-14}.
For convenience, we present in Table~\ref{tab:sig-14bis}, the number of signal events and  significances
at $\sqrt s=14$ TeV, for a reduced  integrated luminosity of $60$ fb$^{-1}$, that could be collected by ATLAS and CMS over about the first year of  LHC running at $\sqrt s=14$ TeV.
In Table~\ref{tab:sig-8}, the corresponding results are shown at  
$\sqrt s=8$ TeV, and integrated luminosity of $60$ fb$^{-1}$, where  a few signal 
events could be detected for negative $C_t$ values, with statistical significances less than 2.

In Table~\ref{tab:sig-14}, one can notice  that, in the $SM$ ($C_f=C_t=1$),  the large integrated luminosities foreseen at the 
high luminosity (HL) LHC project  (a few $10^{3}$fb$^{-1}$) are required in order to measure a 
\tqaaspp signal.

%%%%%%%%%%%%%%%%%%%%%%....Figure...4.
\begin{figure}[t]
\begin{center}
%\vspace*{-2.2cm}
\includegraphics[width=0.6\textwidth]{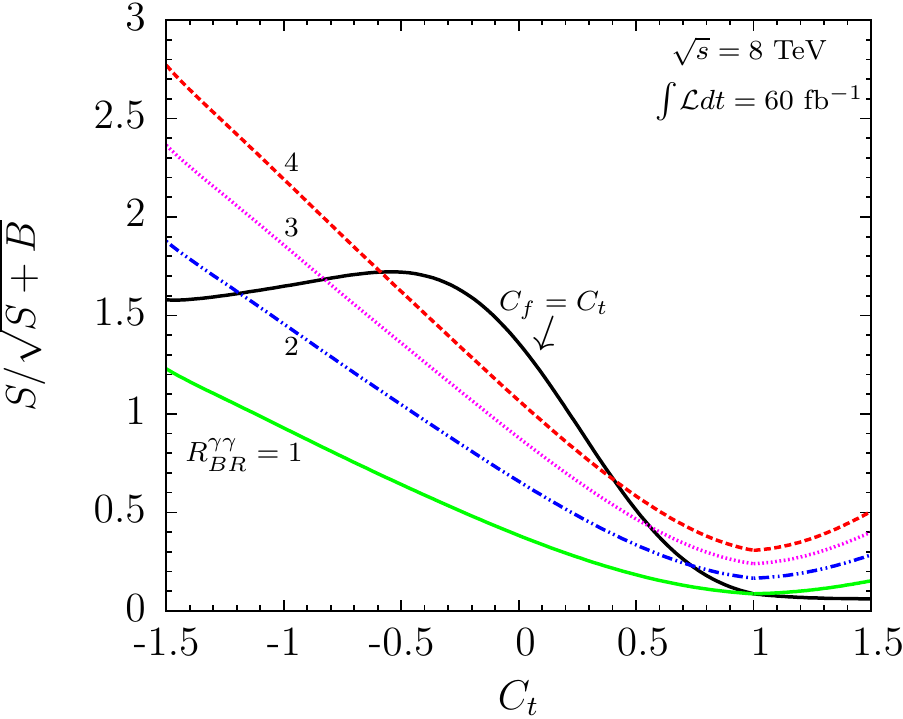}
\vspace{-0.3cm}
%\vspace*{-1cm}
\caption{\small \it Signal significance versus $C_t$. Different assumptions are made for the value of  $R^{\gamma\gamma}_{BR}=$\braas/\braasm. The black solid line represents the 
Yukawa universal-rescaling hypothesis, where $R^{\gamma\gamma}_{BR}$ is just a function of $C_t$, with $C_b=C_{\tau}=C_t$. The remaining (colored) lines refer to the constant $R^{\gamma\gamma}_{BR}=1,2,3,4$ hypothesis.}
\label{fig:signf-8}
\end{center}
\end{figure}
%\vspace{-1.0cm}
%%%%%%%%%%%%%%%%%%%%%%....Figure...5.
\begin{figure}[hp]
\begin{center}
%\vspace*{-2.2cm}
\includegraphics[width=0.6\textwidth]{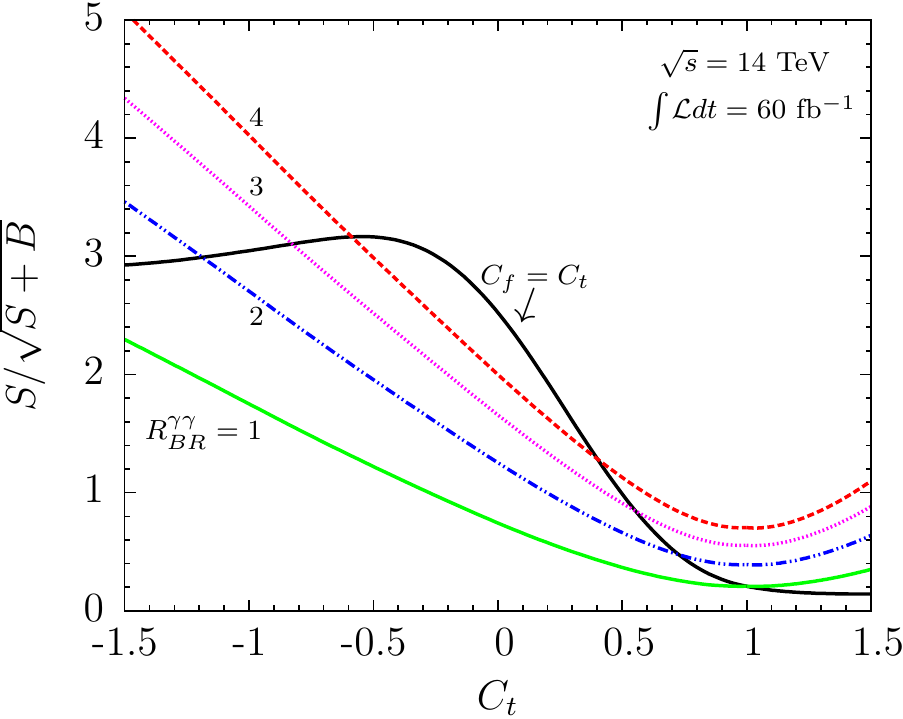}
\vspace{-0.3cm}
%\vspace*{-1cm}
\caption{\small \it Signal significance versus $C_t$. Different assumptions are made for the value of  $R^{\gamma\gamma}_{BR}=$\braas/\braasm. The black solid line represents the 
Yukawa universal-rescaling hypothesis, where $R^{\gamma\gamma}_{BR}$ is just a function of $C_t$, with $C_b=C_{\tau}=C_t$. The remaining (colored) lines refer to the constant $R^{\gamma\gamma}_{BR}=1,2,3,4$ hypothesis.}
\label{fig:signf-14-1}
\end{center}
\end{figure}
%\vspace{-1.0cm}
%===========================================================================================
\begin{figure}[hp]
\begin{center}
%\vspace*{-2.2cm}
\includegraphics[width=0.6\textwidth]{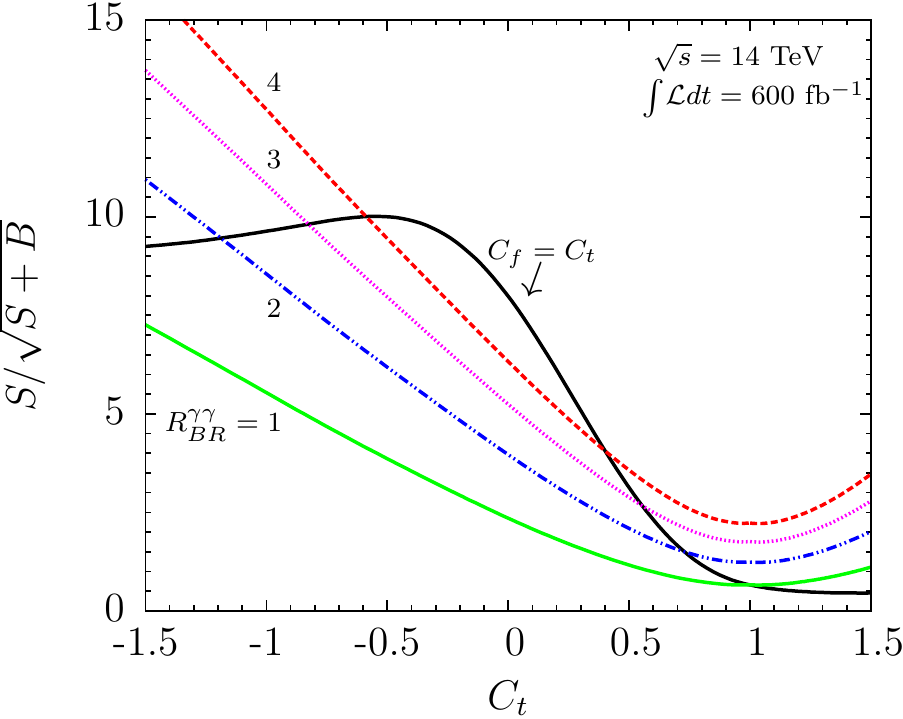}
\vspace{-0.3cm}
%\vspace*{-1cm}
\caption{\small \it Signal significance versus $C_t$. Different assumptions are made for the value of  $R^{\gamma\gamma}_{BR}=$\braas/\braasm. The black solid line represents the 
Yukawa universal-rescaling hypothesis, where $R^{\gamma\gamma}_{BR}$ is just a function of $C_t$, with $C_b=C_{\tau}=C_t$. The remaining (colored) lines refer to the constant $R^{\gamma\gamma}_{BR}=1,2,3,4$ hypothesis.}
\label{fig:signf-14-2}
\end{center}
\end{figure}
%\vspace{-1.0cm}
%===========================================================================================
In Figures \ref{fig:signf-8}--\ref{fig:signf-14-2}, we compare, at different $\sqrt s$ and integrated luminosities, the \tqaaspp signal significances obtained 
in the Yukawa universal-rescaling hypothesis $C_f=C_t$, with the more model-independent framework of fixed values of the ratio $R^{\gamma\gamma}_{BR}=$\braas/\braasm.
In the latter case, the \braas enhancement could arise from a new mechanism beyond the $SM$, that affects only \braas without influencing the \tqhspp production cross section apart from its  $C_t$ dependence. For example, the presence of new heavy physical states could contribute to the \haas width, without affecting the \tqhspp cross section. One can see that 
an enhancement factor $R^{\gamma\gamma}_{BR}\gappeq 2$ is required in order to get at least $3\, \sigma$ significances for $C_t\sim -1$, at $\sqrt s=14$ TeV, and integrated luminosity of $60$ fb$^{-1}$.

\section{Conclusions}
We have analyzed the $t$-channel  \tqaaspp potential for  determining  the relative sign of the $ttH$  and $WWH$ couplings  at the LHC. As previously noted, the \tqhspp production cross section is extremely sensitive to a  sign switch with respect to the $SM$. On the other hand, the actual potential of the single-top plus Higgs associated production depends on the additional theoretical assumptions on \braa.
We have made a parton-level simulation of the \haas decay signal for the \tqhspp channel, and studied the corresponding main irreducible backgrounds with a quite conservative set of  selection cuts on the kinematics of the final particles. We have found that the first year of the LHC running at 14 TeV could be sufficient, if $C_f=C_t\lappeq 0$, to have a considerable signal event number with moderate  background. In particular, an integrated luminosity of $60$ fb$^{-1}$
would give about 10 signal events versus less than 0.3 background events over all the negative range $-1.5\lappeq C_t\lappeq 0$. This  is to be confronted with the result corresponding to the $SM$ parameter setup, that would require the integrated luminosities of the HL-LHC in order to reach an observable event statistics.

We then  urge the LHC experimental groups to consider the single-top plus Higgs associated production
\tqas through a full-simulation analysis.
We also leave to further work the assessment of the potential of the \tqhspp process as a probe of an anomalous top Yukawa behavior by means of the Higgs decays other than \haa.

\vspace{1cm}
\vbox{
\noindent{\bf Acknowledgments}  \\
\vspace{-0.2cm}
\noindent \\
We thank Fabio Maltoni for pointing out to us the single-top plus Higgs production sensitivity to anomalous top Yukawa couplings.
Discussions with  Aleksandr Azatov, Roberto Contino, Kirtiman Ghosh, Pradipta Ghosh, 
Satyanarayan
Mukhopadhyay, and Saurabh Niyogi are thankfully acknowledged. We also thank 
Gennaro Corcella, Leandro Nisati,  and Fulvio Piccinini for advice on  event simulation, and  
the RECAPP, Harish-Chandra
Research Institute, for providing some extra cluster computing resources.
E.G. would like to thank the PH-TH division of CERN for its kind 
hospitality during the preparation of this work.
This work was supported by the ESF grant MTT60, by the recurrent financing SF0690030s09 project and by  the European Union through the European Regional Development Fund.
}

\end{document}